\documentclass{iau} 
\usepackage{graphicx}

\title[Quantitative spectroscopy in the Galaxy] 
{Highly accurate quantitative spectroscopy  \\ of massive stars in the Galaxy}

\author[M.-F. Nieva and N. Przybilla]   
{Mar\'ia-Fernanda Nieva and Norbert Przybilla}

\affiliation{Institut f\"ur Astro- und Teilchenphysik, Universit\"at
Innsbruck, \\ Technikerstrasse 25/8,
6020 Innsbruck, Austria \\ email: {\tt maria-fernanda.nieva@uibk.ac.at} \\[\affilskip]
}

\pubyear{2017}
\volume{329}  
\jname{The lives and death-throes of massive stars}
\editors{J.J. Eldridge, L. McClelland, L. Xiao \& J. Bray, eds.}
\begin{document}

\maketitle

\begin{abstract}

Achieving high accuracy and precision in stellar parameter and chemical composition 
determinations is challenging in massive star spectroscopy. 
On one hand, the target selection for an unbiased sample build-up is
complicated by several types of peculiarities that can occur
in individual objects. On the other hand, composite spectra are often not recognized 
as such even at medium-high spectral resolution and typical signal-to-noise ratios, 
despite multiplicity among massive stars is widespread. In particular,
surveys that produce large amounts of automatically reduced data are prone
to oversight of details that turn hazardous for the analysis with
techniques that have been developed for a set of standard
assumptions applicable to a spectrum of a single star. Much larger systematic 
errors than anticipated may therefore result because of the unrecognized true 
nature of the investigated objects, or much smaller sample sizes of
objects for the analysis than initially planned, if recognized. More factors to
be taken care of are the multiple steps from the choice of instrument
over the details of the data reduction chain to the choice of
modelling code, input data, analysis technique and the selection of
the spectral lines to be analyzed. Only when avoiding all the possible
pitfalls, a precise and accurate characterization of the stars in
terms of fundamental parameters and chemical fingerprints can be
achieved that form the basis for further investigations regarding e.g. 
stellar structure and evolution or the chemical evolution of the Galaxy.
The scope of the present work is to provide the massive star and also other astrophysical 
communities with
criteria to evaluate the quality of spectroscopic investigations of
massive stars before interpreting them in a broader context. The
discussion is guided by our experiences made in the course of over a
decade of studies of massive star spectroscopy ranging from the {\em simplest}
single objects to multiple systems.

\keywords{stars: parameters, stars: abundances}
\end{abstract}

\firstsection 
\section{Introduction}
We focus our discussion on stars with masses between $\sim$20 and $\sim$6 solar masses.
From the spectral characteristics, this corresponds to the latest O
stars to early B stars on the Main Sequence, the most numerous but
also the {\em simplest} massive stars to be studied.
In contrast to earlier spectral types, supergiants or Be stars, they
are not significantly affected by mass outflow or complex geometries.
And, in comparison to many mid/late B-type stars, they do not present signatures of 
atomic diffusion in their atmospheres.
Many model atmospheres and analysis methods for massive star spectra
have therefore been optimized to study these simple objects. The
typical assumptions are that they are single, they have stationary, hydrostatic and chemically
homogeneous atmospheres, are described well by plane-parallel geometry
and (non-)local thermodynamic equilibrium.

However, in order to address the topic of massive stars in its
entirety, one has to analyze earlier O stars as well, OB supergiants,
Be and late B-type stars, chemically peculiar, magnetic, pulsating, fast-rotating stars, and
also pre-main-sequence stars. The large variety of massive stars challenges the basic assumptions 
of standard atmospheric models and spectral analysis techniques that have been often used 
to estimate their fundamental parameters and chemical composition. One
has to consider e.g.~(inhomogeneous) hydrodynamic outflows, oblateness of the stars because of high
rotational velocities and the presence of an accretion/decretion disk
surrounding them (i.e. non-spherical geometries), 
the effects of spots on stellar surfaces due strong magnetic fields,
elemental abundance stratification due to diffusion, deformation of spectral lines
due to stellar pulsations, etc. Practically all these topics have been
addressed by dedicated studies. However, a unique tool to
(automatically) analyze spectra formed under such a variety of conditions is not implemented 
currently. The limiting factor is the large number of physical
assumptions to be made for the general case, preventing an efficient
calculation of models covering the entire parameter space.

A more feasible approach to reliably study each kind of star is via
tailored model atmospheres and analysis techniques that incorporates the most realistic 
physical background for each particular case. Dedicated spectral grids can then be computed 
for each kind of object when the range in all relevant parameters is known.
The final scope is to derive the stars' atmospheric parameters to reliably place them in a 
Kiel diagram (surface gravity $\log g$
vs.~effective temperature $T_\mathrm{eff}$). In cases where fundamental parameters can also be 
derived (with additional information, like e.g.~parallaxes),
the objects can be placed in a Hertzsprung-Russell diagram, and basic
relationships between fundamental stellar parameters can be checked.
Related to the atmospheric and fundamental parameters is the determination of 
the stars' spectroscopic distances,
that -- when compared with other independent distance indicators -- allow us to constrain 
the accuracy of the stellar parameters.
Spectral energy distributions from the UV to the IR allows us to derive the extinction and 
reddening for a particular star, and to cross-check for it's global
energy output.

A subsequent step after the stellar parameter determination is the
chemical analysis of trace elements, which also differs among the various
kind of stars because of e.g., temperature constraints, where lines of different 
species/ionization stages vary in their strength, different blends appear
in different parts of the spectrum, or some lines can turn from
absorption to emission. The projected rotational velocity will
also determine which lines can be analyzed in isolation and which blends can be 
consistently taken into account.
A detailed chemical analysis can yield on accurate elemental abundances when four basic 
conditions are met:
i) the stellar atmospheric models are applicable to the objects, ii) the spectra have a good quality 
(spectral lines should be resolved and measurable at good S/N), iii) the analysis methodology takes 
into account a good selection of spectral lines that can be well reproduced by the model spectra 
and iv) the atmospheric (and when possible fundamental)
parameters have been consistently derived.

This manuscript is biased towards our experience on the spectral modeling and analysis of normal and
peculiar massive single and recently also multiple stars. It discusses some successes but also 
challenges to current modeling capabilities.
It is intended as a guideline for colleagues working in other areas to
{\em assess} the accuracy and precision of published fundamental parameters and chemical 
abundances for massive stars, before they are interpreted in various astrophysical contexts.
A special emphasis is put on the single OB stars on the Main Sequence, because they allow us to 
derive most parameters
and chemical abundances at highest accuracy and precision, and therefore, we can consider 
them as reference objects.
Shorter descriptions are dedicated to particular classes of more complex objects for which we have
recently extended our spectral analyses.

\section{Late-O and early B-type on the Main Sequence: the {\em simplest} stars}

\begin{figure}[ht!]
  \centering
  \begin{minipage}{0.48\textwidth}
    \includegraphics[width=\textwidth]{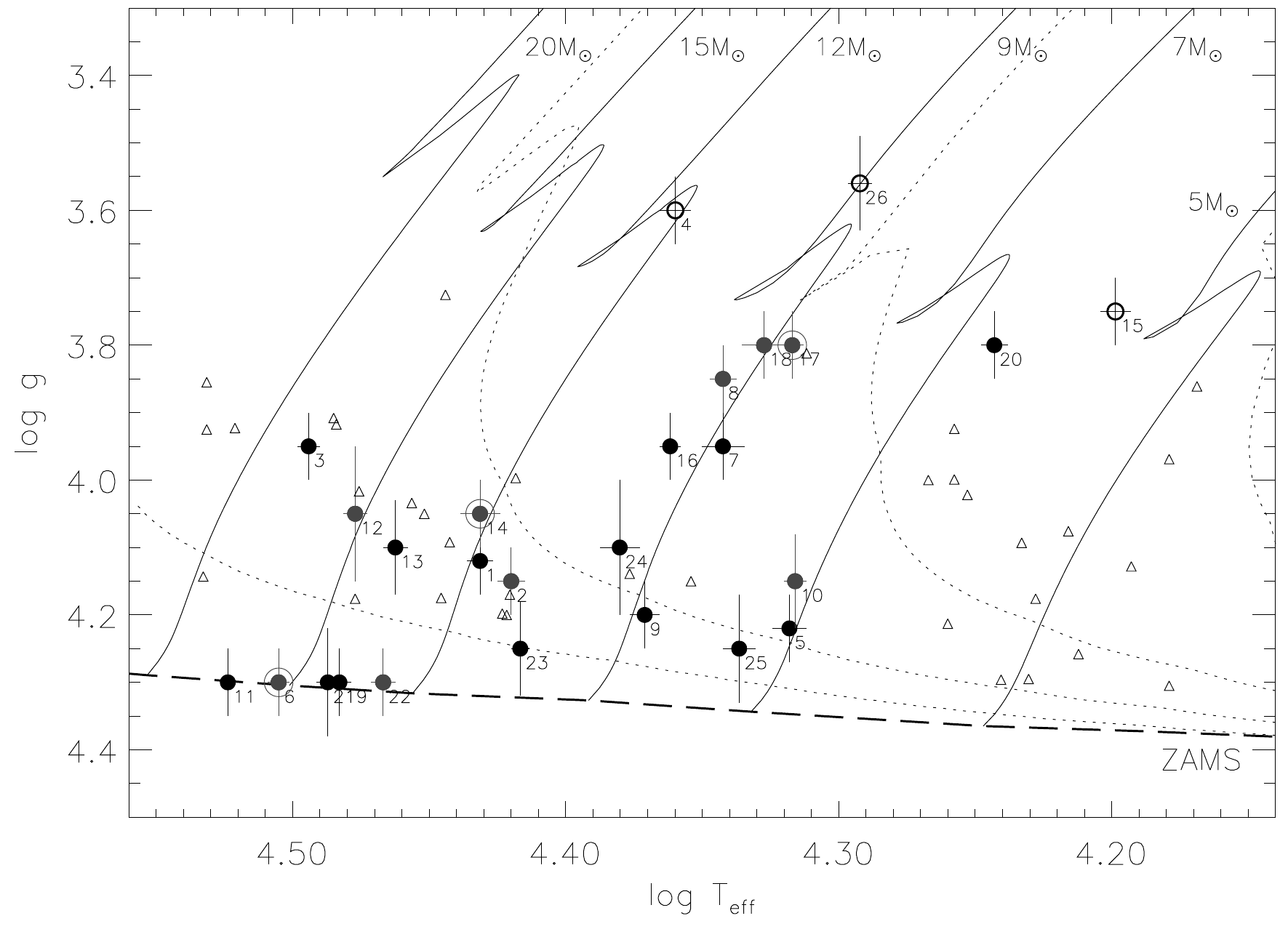}
    \caption{$T_\mathrm{eff}$ and $\log g$ of a sample of early B-type stars
    on the Main Sequence (black dots).
    Open thick circles are objects beyond core H-exhaustion. Wide circles
    surrounding the dots mark magnetic stars.
    Data from double-lined detached eclipsing binaries are shown as small triangles.
    Evolutionary tracks and isochrones
    from \cite{2012A&A...537A.146E} are shown. See NP12 and NP14 for details.}
    \label{fig1}
  \end{minipage}
  \quad
  \begin{minipage}{0.48\textwidth}
    \includegraphics[width=\textwidth]{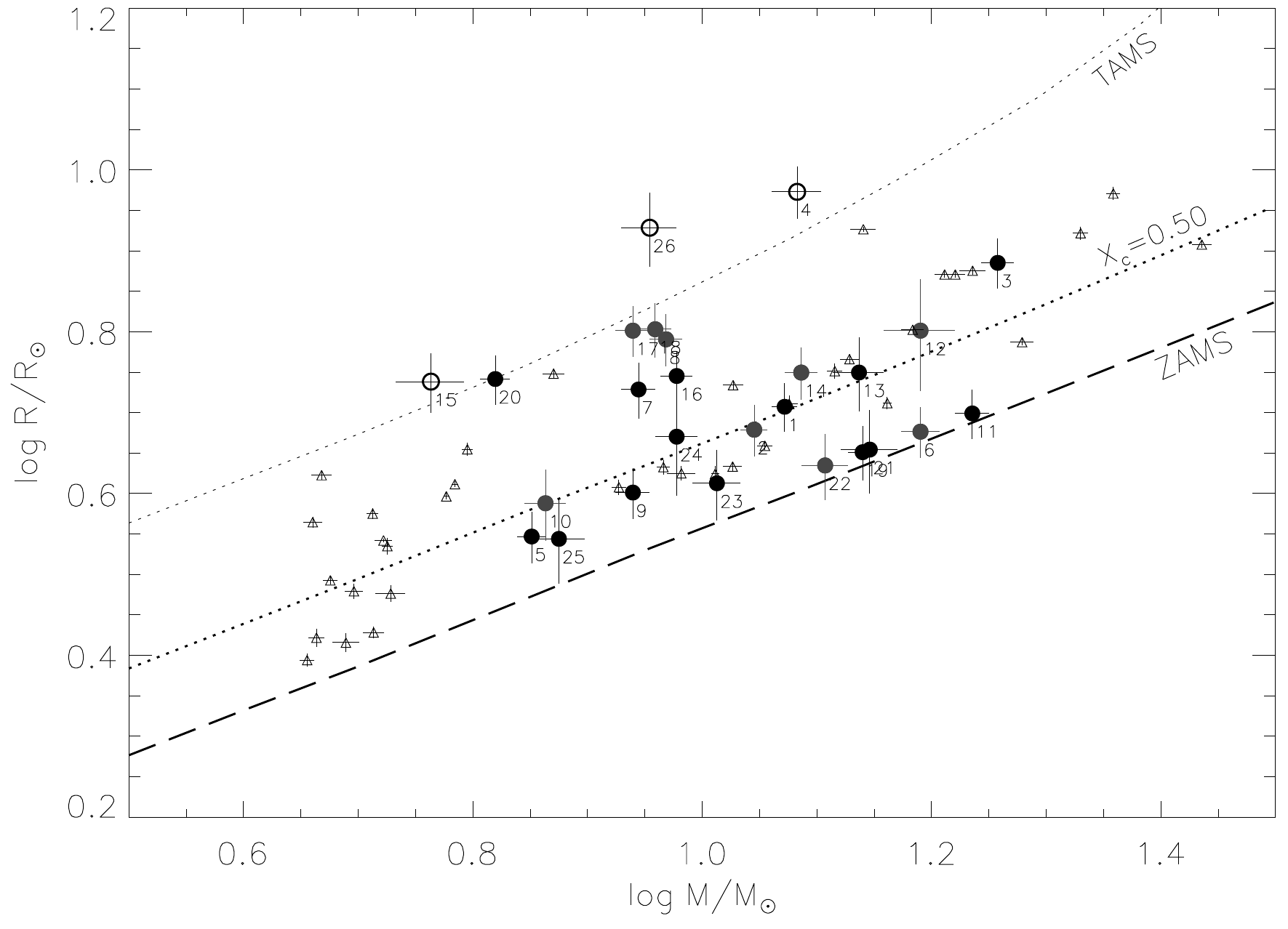}
    \caption{Mass-radius relationship for the sample stars in Fig.~\ref{fig1} with the same symbol encoding.
    Abscissa values are evolutionary masses. The ZAMS and two additional loci for 50\% core-H depletion and for the TAMS
    are indicated by the thick/thin-dotted lines from the stellar evolution code as in Fig.~\ref{fig1}.
    Error bars are shown also for the detached eclipsing binaries. See NP14 for details.}
    \label{fig2}
  \end{minipage}
\end{figure}

\begin{figure}[ht!]
  \centering
  \begin{minipage}{0.48\textwidth}
    \includegraphics[width=\textwidth]{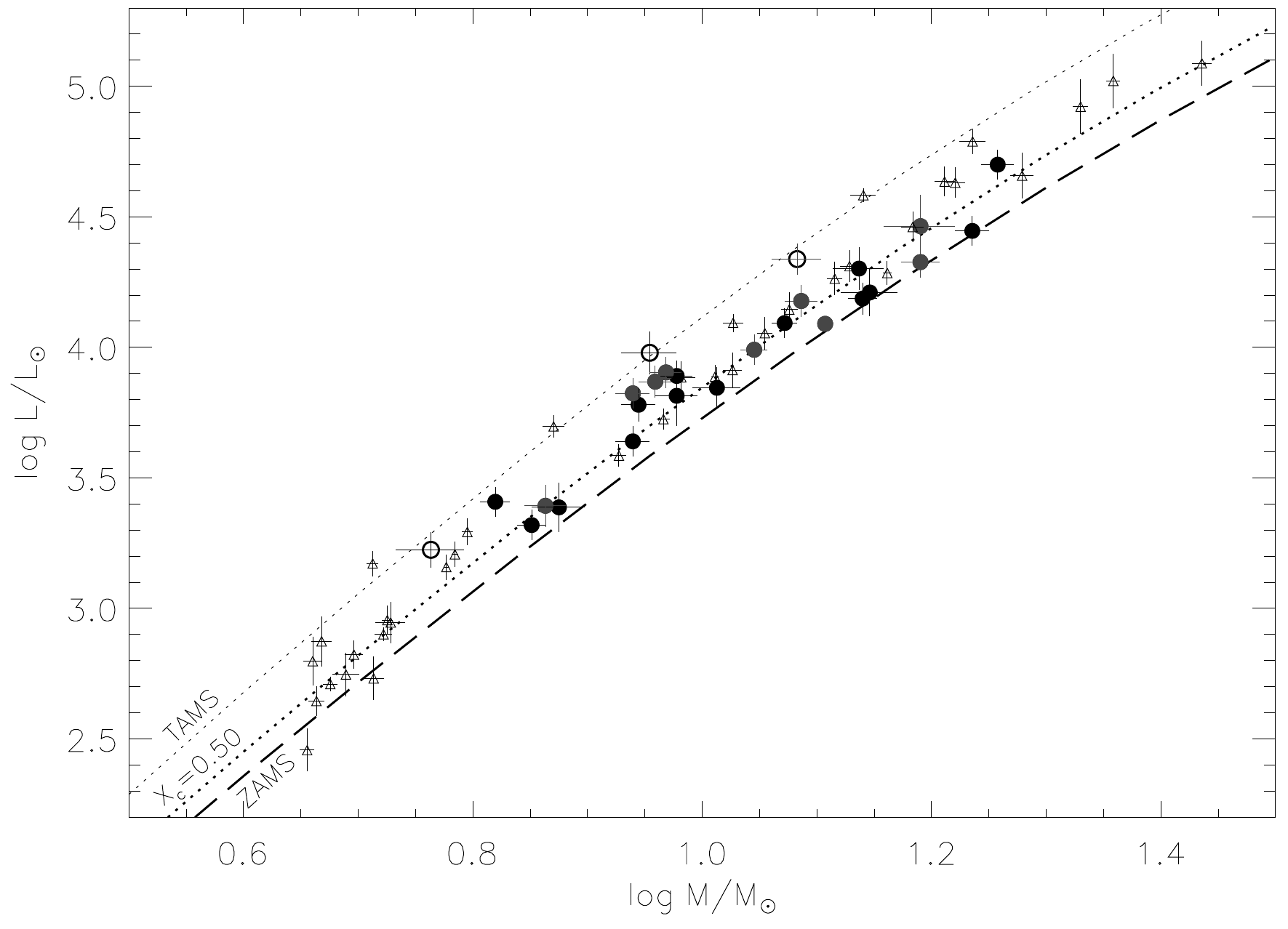}
    \caption{Mass-luminosity relationship for the sample stars in Fig.~\ref{fig1}. Symbol and loci
    encoding are the same as in Figs.~\ref{fig1} and \ref{fig2} (NP14).}
    \label{fig3}
  \end{minipage}
  \quad
  \begin{minipage}{0.48\textwidth}
    \includegraphics[width=\textwidth]{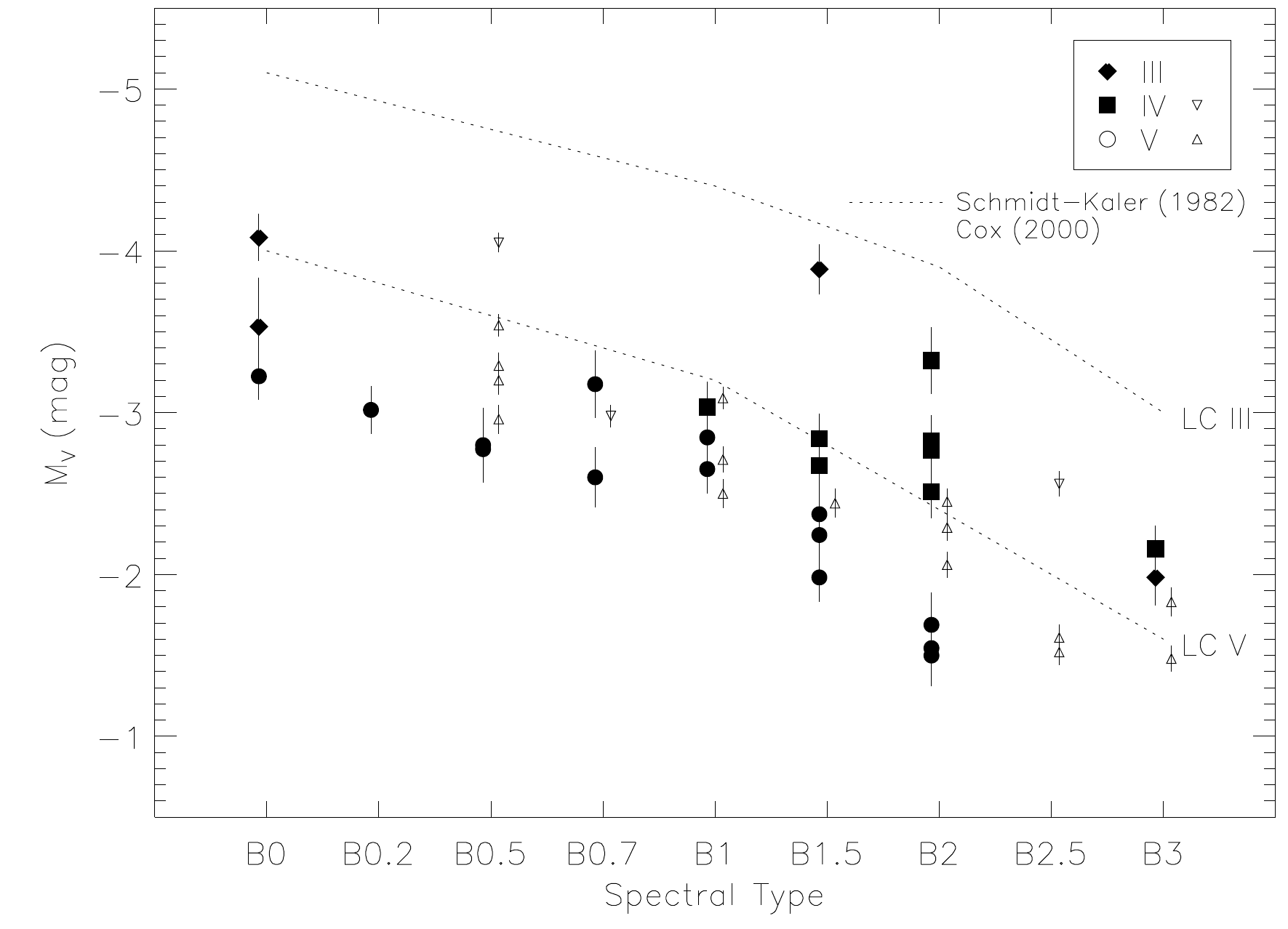}
    \caption{Absolute visual magnitudes of the sample stars in Fig.~\ref{fig1}, encoding their luminosity class
    according to the legend, vs. spectral type. Older reference values are also indicated (NP14).}
    \label{fig4}
  \end{minipage}
\end{figure}

For over a decade, we have improved the spectral modeling and analysis technique for stars with spectral classes
from O9 to B2 and luminosity classes V to III -- the so-called OB subgroup because of their similar spectral characteristics.
Given that standard stellar model atmospheres like e.g. {\sc Atlas9} \cite{1993KurCD..13.....K}
meet most requirements to reproduce their
atmospheric structures well (\cite[Nieva \& Przybilla 2007]{2007A&A...467..295N}), our efforts were invested into realistic
level population and line-formation computations in
non-LTE by building and testing different configurations of new input atomic data to provide with robust model atoms
for different elements (see \cite[Przybilla et al. 2016]{2016A&A...587A...7P} for updated references).
The codes used for the non-LTE level population and line-formation calculations are {\sc Detail} \cite{Giddings1981Auth}
and {\sc Surface} \cite{1985ButlerGiddings}\footnote{For this type of
stars, LTE atmospheric structures computed
with {\sc Atlas9} and stellar fluxes computed with {\sc Atlas9} and {\sc Detail} are practically
identical to those calculated with {\sc Tlusty}
\cite{1995ApJ...439..875H} in non-LTE, see \cite{2007A&A...467..295N}
for a detailed discussion.}.
Our new spectral modeling in combination with a self-consistent spectral analysis that accounts for multiple ionization
equilibria, applied to high-quality observations, resulted in a drastic minimization of stellar parameter and chemical abundance
uncertainties, particularly reducing several systematic effects previously unaccounted for.
A comprehensive study and first
applications of our work on single and normal early B-type dwarfs and giants were discussed by \cite[Nieva \& Przybilla (2012, 2014, hereafter NP12 and NP14)]{2012A&A...539A.143N,2014A&A...566A...7N} and \cite{2013A&A...550A..26N}.
These studies allowed us to put constraints not only on their atmospheric parameters
and chemical composition at high precision and accuracy, but also to derive spectroscopically other stellar parameters like
radius, luminosity, mass and to explore whether theoretical relations between them hold using our observationally
derived parameters.
For such stars, uncertainties in effective temperature, surface gravities and chemical abundances as low as
1-2\%, 15\% and 25\%, respectively (NP12) and in evolutionary masses, radii and luminosities
 better than 5\%, 10\%, 20\%, respectively
and in absolute visual and bolometric magnitudes
lower than 0.20 mag (NP14) are achieved.  Moreover, practically a perfect match
between the observed and the computed spectrum per star for one set of parameters confirm the robustness of models and analysis.
Figure~\ref{fig1} shows a precise location of a sample of single early B-type stars (NP12) in the Kiel diagram
in comparison to detached eclipsing binaries (DEBs). Figure~\ref{fig2} shows their mass-radius relation and
Fig.~\ref{fig3}
their mass-luminosity relation (NP14), indicating that the derived stellar parameters lie within theoretically expected values and agree with more accurate results from DEBs.
The empirical relation between absolute magnitudes and spectral types in NP14 show an offset with respect to
classical older calibrations from the literature \cite{2000asqu.book.....C}. The latter are often
adopted, affecting the computation of e.g. the stellar luminosities.
The level of accuracy reached for this kind of objects can hardly be reproduced for more
complex stars or systems.
We therefore use these results as references for further studies.

\section{BA and OB supergiants}

Supergiants of spectral types late-B and early-A are descendants of OB stars on the Main Sequence.
Their atmospheres can be well modeled with {\sc Atlas9}, {\sc Detail} and {\sc Surface},
as described in \cite{2006A&A...445.1099P}. These objects also have the advantage that they 
outshine practically any companion in the visual, facilitating to treat them as single stars.
The spectral analysis of early-O stars and OB supergiants is beyond the modeling capability 
of this hydrostatic approach, however
other robust hydrodynamic non-LTE codes that treat simultaneously the
spherically symmetric stellar atmospheric structure and the stellar wind like {\sc Cmfgen} \cite{1998ApJ...496..407H}
and {\sc Fastwind} \cite{2005A&A...435..669P} are suited for their study.
\section{Fast-rotating stars}\label{fast}
We encounter several limitations in the analysis of intrinsically fast-rotating stars.
If the star rotates so fast that its shape is affected (oblate), the assumptions of plane-parallel or
even spherical symmetry
are no longer valid on the global scale. Also the temperature will vary between the poles and the equator. Depending on the star's inclination, we observe spectral lines formed at
different atmospheric conditions, with different local temperatures. This is clearly evident in cases of fast-rotators seen pole-on,
where the lines are sharp, but nevertheless it is difficult to fit the whole spectrum with one temperature only and
the assumption of establishing ionization balance for all available species simultaneously is no longer valid.
If the projected rotational velocity is large, only a few spectral lines can be analyzed and many blends have to be accounted for self-consistently. If the S/N ratio is not high enough, the definition of the line continuum is one of the largest sources of systematics
because the spectral lines get smeared out and appear weaker (see \cite{2005ApJ...633..899K} for a discussion).
The analysis of fast-rotating stars is challenging and the limitations have to be established on a case-by-case basis.
One of the most difficult tasks, when the stars have large projected rotational velocities, is identifying systematic asymmetries
that can be caused by a companion in a spectroscopic binary (or multiple) system.
Stellar parameters and chemical abundances derived from these kind of objects should be treated extremely carefully,
because they are prone to large systematic errors.  \cite[Maeder et al. (2014)]{2014A&A...565A..39M}
 provide in their Table~1
 an example of a re-assessment of a sub-sample of objects from the {\sc Flames} Massive Star Survey studied by
\cite{2009A&A...496..841H}, resulting in the identification of stars with previously unnoticed double or asymmetric spectral
lines, which cause their natural exclusion from the interpretation of results because they were analyzed with standard
techniques.

\section{Be stars}
Be stars are main-sequence or subgiant stars of spectral type B characterized by Balmer emission (e.g., H$\alpha$, but also other Balmer and
eventually metal lines) that originates in a circumstellar disk.
As most Be-stars are fast-rotators, the same limitations exposed in Sect.~\ref{fast} apply to their
analysis. Additionally, the clear signatures of lines formed
in the disk cannot be reproduced self-consistently with any stellar atmosphere code at present.
However, there are cases where the photospheric lines can be
successfully reproduced, constraining reliably the stellar parameters and even metal abundances. An exploration of the extent of
applicability of our spectral models and methods to such stars is ongoing.

\section{Mid- and late B-type stars}
Mid- and late B-type stars present fewer metal lines and less elements with different ionization stages traced in the spectra
(1 or 2), in contrast to hotter stars (4 to 6). In many cases, in addition, the stars have a large
projected rotational velocity, which can turn their analysis even more challenging. The identification 
of a companion forming a spectroscopic binary system, by resolving line-asymmetries is also a challenge 
for many of these stars (Zwintz et al., subm.).
Many objects show chemical peculiarities due to atmospheric diffusion, others present normal
chemical abundances. 

\begin{figure}[ht!]
  \centering
  \begin{minipage}{0.48\textwidth}
    \includegraphics[width=\textwidth]{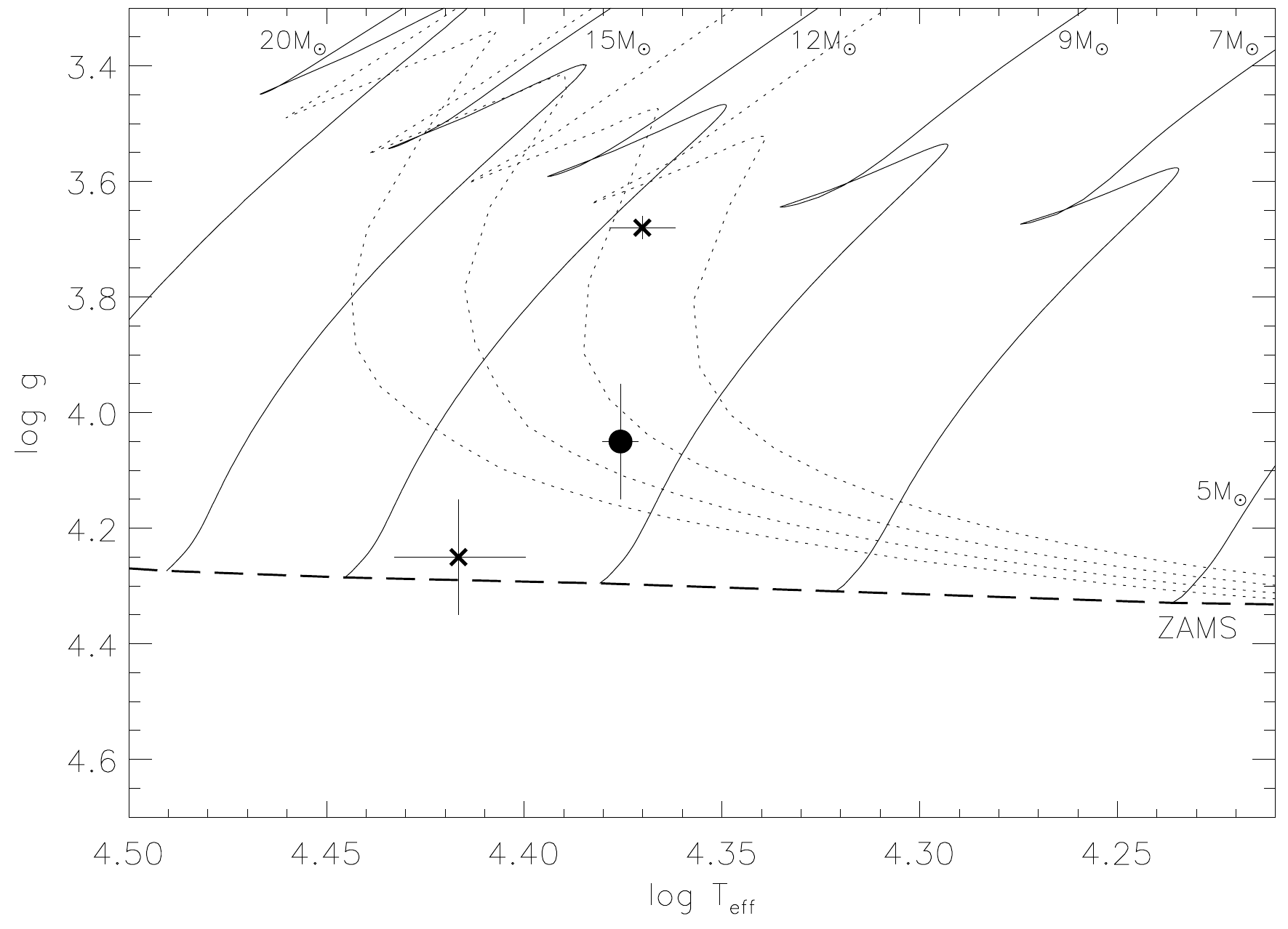}
    \caption{CPD$-$57\,3509 in the Kiel diagram. Black dot: result from our self-consistent analysis.
    St. Andrew’s crosses: results discussed in the literature at higher and lower lower gravity
    assuming solar helium abundance. Adapted from \cite{2016A&A...587A...7P}.
     }
    \label{fig5}
  \end{minipage}
  \quad
  \begin{minipage}{0.48\textwidth}
    \includegraphics[width=\textwidth]{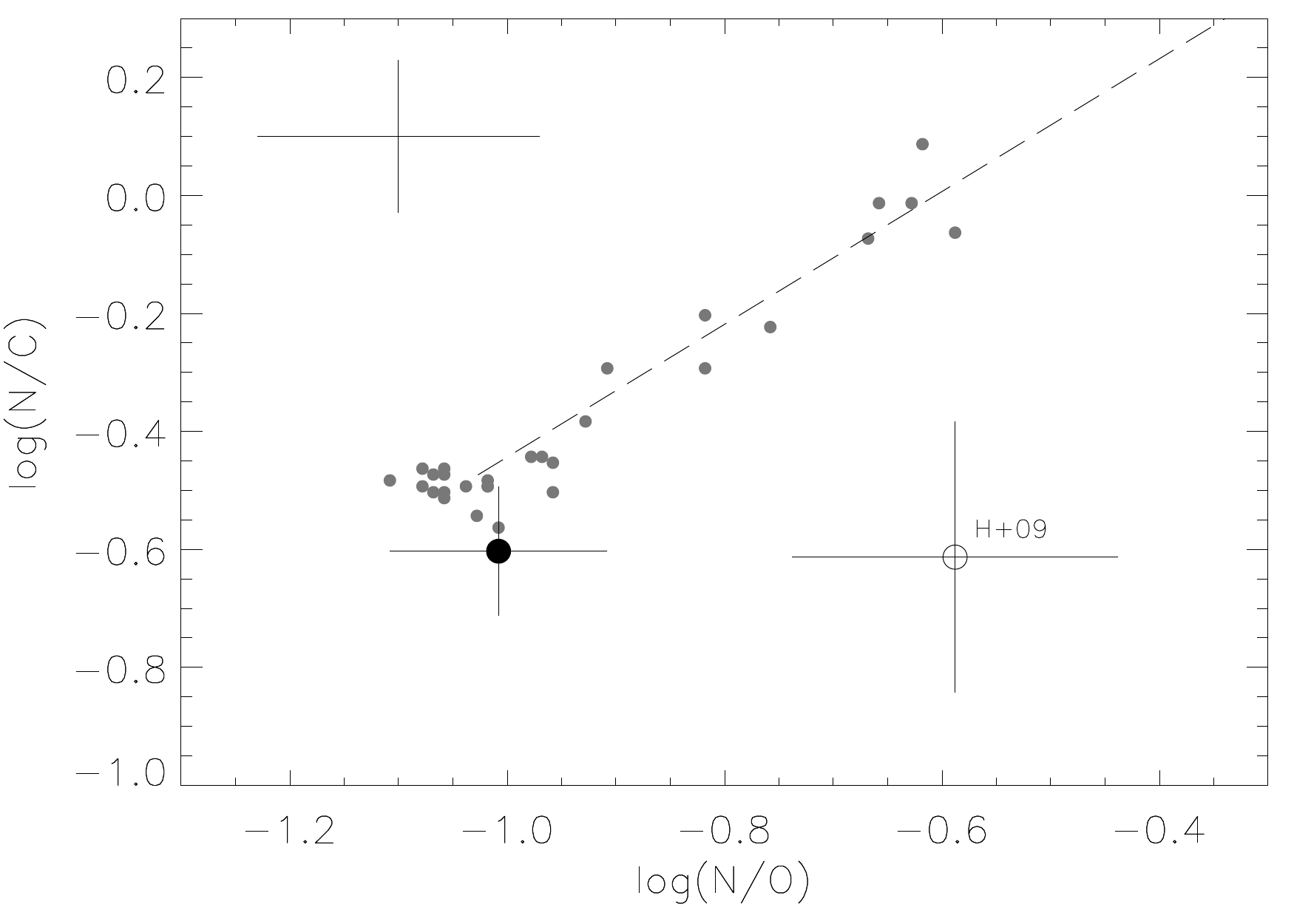}
    \caption{CPD$-$57\,3509 in the N/O-N/C diagram (by mass). Black dot: \cite{2016A&A...587A...7P}.
    Open circle:  \cite{2009A&A...496..841H}. Grey dots: early B-type stars from NP12.
    Dashed line: analytical approximation to the nuclear path for CN-cycled material using initial values from NP12.}
    \label{fig6}
  \end{minipage}
\end{figure}

\section{Chemically peculiar, pulsating and magnetic stars}
Chemically peculiar stars are stars with distinctly unusual metal abundances in their surface layers.
The stars present selective diffusion of different elements in their atmospheres causing some elements 
to show higher or lower abundances in their outer layers.
Helium-strong stars constitute the hottest and most massive chemically peculiar stars of
the upper Main Sequence.
The chemical abundances and metallicities of such stars should be accounted for iteratively in 
every step of the
analysis by computing dedicated model spectra, because their peculiar abundances are not considered 
in pre-computed grids of model spectra.
An example of the consequence of not taking the peculiar abundances into account in the 
stellar parameter and chemical composition determination,
in contrast to a self-consistent analysis is shown in Figs.~\ref{fig5} and~\ref{fig6}, adapted from
\cite{2016A&A...587A...7P}.
Analyses assuming solar helium abundances, in contrast to the enhanced helium abundance present
in the star, can result in different atmospheric parameters, as seen in 
Fig.~\ref{fig5} and consequently in different chemical abundances, as shown in Fig.~\ref{fig6}.
The same methodology used by \cite{2016A&A...587A...7P} was consequently applied to other 
He-strong stars by \cite{gonzalez17}, \cite{castro17}, Briquet et al. (in prep.).

Other less-massive chemically peculiar stars (mid- to late B-type) like He-weak or HgMn stars may pose an extra challenge in their spectral analysis caused
by elemental abundance stratification in their atmospheres. Clear signatures of He abundance stratification are noticed e.g.
in $\kappa$\,Cancri, even when the modeled spectra are computed in non-LTE, whereas oxygen seems not to be stratified in non-LTE, therefore the
ionization balance using lines of O\,{\sc i} and O\,{\sc ii} formed in different depths in the atmosphere is met when considering
non-LTE line formation \cite{2014A&A...572A.112M}.

Strongly-pulsating stars present spectral lines with noticeable asymmetries. As model atmosphere codes usually do not include
modes of stellar pulsations in their computations, such asymmetries
cannot be reproduced typically, but see \cite{2016A&A...591L...6I}.
We notice, however, that some spectral lines are more symmetric because they are formed
in regions where the pulsations affect the atmosphere less. In a case-by-case study, it is possible 
to derive atmospheric parameters through the analysis of the most symmetric lines. An example
can be found in \cite{2011AA...527A.112B}. Note, however, the difference in $\log g$ resulting
from the spectroscopic and from the asteroseismic analysis in that case, which should be further 
investigated.
We were also able to reproduce the spectra of several magnetic stars, with magnetic field strengths ranging from very
weak to very strong. Some examples are discussed in \cite{2015A&A...574A..20F}, \cite{2016A&A...587A...7P},
 \cite{gonzalez17}, \cite{castro17}, Briquet et al. (in prep.), within the "B fields in OB stars" (BOB) collaboration.

\begin{figure}[ht!]
  \centering
    \includegraphics[width=\textwidth]{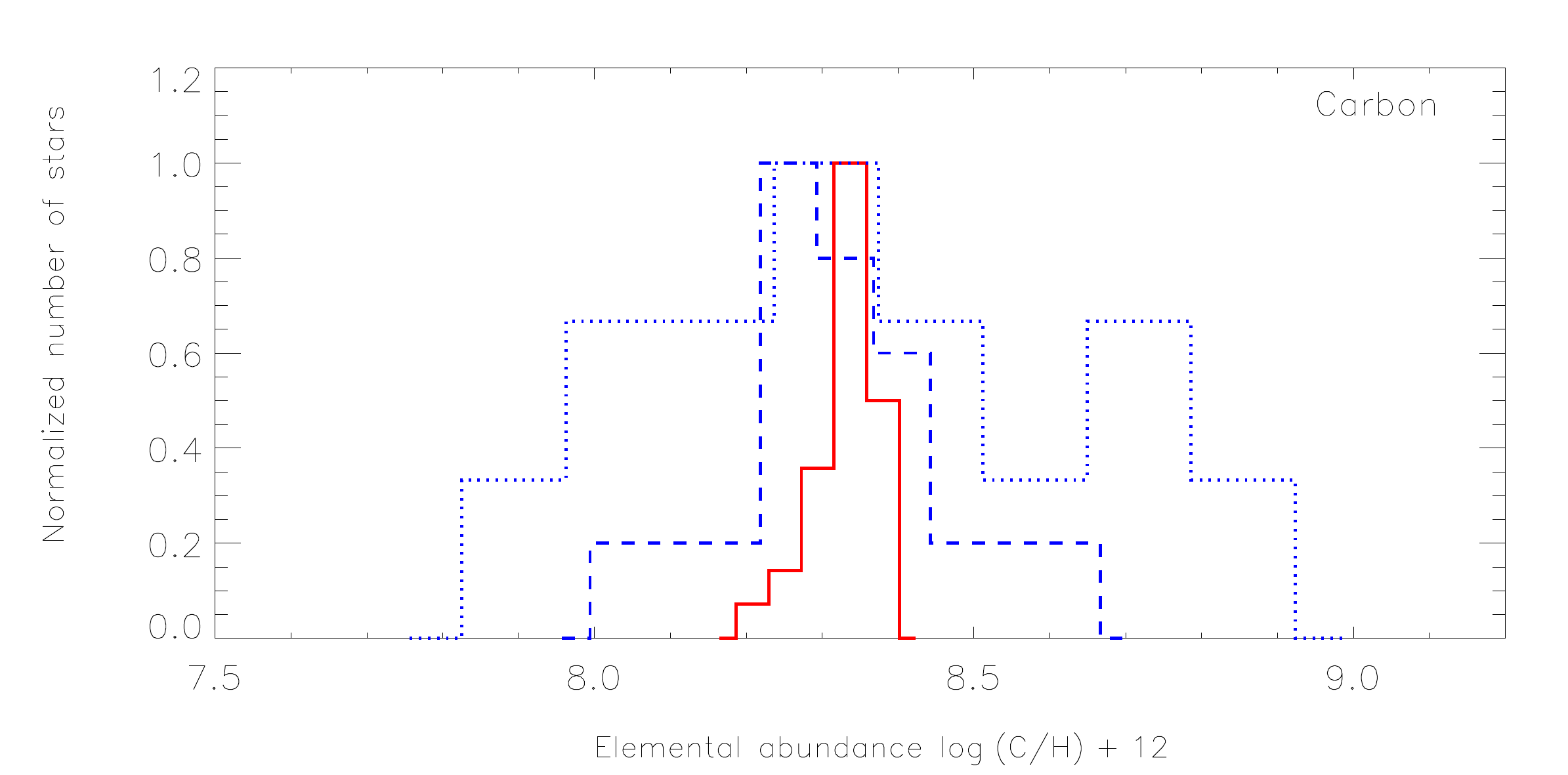}
    \caption{Present-day carbon abundances derived from early B-type stars in the Galaxy.
Blue dashed line: apparently normal single stars. Blue dotted line: previously unnoticed 
SB2 candidates. The stars are distributed in the Galactic disc covering 4 to 15 kpc Galactocentric distance.
For comparison, in red full line are carbon abundances of normal single early B-type stars in the 
solar neighborhood as derived by NP12. 
    }
    \label{lastfig}
\end{figure}

\section{Multiple systems}

The spectral line fitting of the triple system HD\,164492C in the
Trifid Nebula, which is
formed by a close spectroscopic binary composed by a normal early B-type star and a
chemically peculiar late B-type star and a tertiary He-strong star,
has been successfully realized by us in \cite{gonzalez17}.
The number of parameters involved in the spectral fitting is much larger than in the case of single stars.
The challenge of the analysis consists in the constraint that the spectroscopic solution should provide
consistent ages, spectroscopic distances, mass ratio of the primary system
and flux scaling factors. And additionally, because of the chemically peculiar composition
of the He-strong star, dedicated grids of model spectra have to be computed to derive parameter
combinations for $T_\mathrm{eff}$, $\log g$, microturbulence, projected rotational velocity,
macroturbulence, helium abundance and individual metal abundances per star.
Within the whole procedure, several parameters are interrelated with each other, therefore
the analysis is intensively iterative.
In the particular case of HD\,164492C, the Balmer lines do not allow to distinguish the contribution of the
two brighter components,
but the He {\sc i} lines and the stronger Si {\sc iii} lines show the contribution of
the faster-rotator He-strong star on the spectral line wings.
The parameters of the faintest star (with a flux contribution of about
4-5$\%$) cannot be determined from 
the composite spectrum. Instead, the disentangled spectrum can provide us
with some constraints on stellar parameters. 
This kind of analysis is feasible to be done when time series spectra
are available, which is time-consuming for faint objects.

\section{Present-day carbon abundances in the Galactic disk as an application of accurate quantitative 
spectroscopy in the Galaxy}

We have analyzed new high-resolution spectra taken with UVES@VLT of early B-type stars distributed along 
the Galactic disk covering Galactocentric distances ranging from 4 to 15 kpc. 
The sample was selected from the literature, and the high-quality data
allowed us to identify new SB2 candidates.
Preliminary carbon abundances are presented in Fig.~\ref{lastfig}. The whole sample has been analyzed 
assuming them as single stars, as it was done in previous work. However, if we separate stars with 
previously unnoticed SB2 signatures (blue dotted line, where the
chosen analysis approach is inappropriate),
the carbon abundance distribution is much broader than the sample of apparently single stars 
(blue dashed line), which still should present an intrinsic spread due
to effects of the Galactic abundance gradient.
Carbon abundances of apparently single early B-type stars in the solar neighborhood 
(red full line) as derived in NP12, are also plotted for comparison. 

This is a clear case where -- besides improvements in the spectral
modeling and analysis technique -- the data quality (spectral resolution, S/N ratio and 
wavelength coverage) is decisive for avoiding systematic errors. In
this case, the source of systematics lies in the previously unrecognized composite nature 
of some spectra, which need to be excluded from the analysis based on standard assumptions.

\acknowledgements{MFN acknowledges support by the Austrian Science Fund through the Lise
Meitner program N-1868-NBL.}

\end{document}